\documentclass[aps,showpacs,twocolumn,superscriptaddres,nofootinbib]{revtex4-1}

\usepackage{graphicx}

\usepackage{epstopdf}
\usepackage{color, amsmath, bm}
\usepackage[dvipsnames]{xcolor}
\usepackage[hidelinks]{hyperref}
\usepackage{natbib}
\usepackage{circuitikz}

\usepackage{amsfonts}   
\usepackage{physics} 

\usepackage{subcaption}
\usepackage{caption}
\usepackage{float}

\usepackage[normalem]{ulem}

\usepackage{mathtools}
\mathtoolsset{showonlyrefs}

\begin{document}

\title{Dissipative Effects in Transmission Line Analogues of Hawking Radiation}

\author{Elián Urtubey}
\email{e.urtubey@df.uba.ar}
\author{Fernando C. Lombardo}
\email{lombardo@df.uba.ar}
\author{Paula I. Villar}
\email{paula@df.uba.ar}
\affiliation{Departamento de Física Juan José Giambiagi and IFIBA UBA-CONICET, Facultad de Ciencias Exactas y Naturales, Ciudad Universitaria, Pabellón I, 1428 Buenos Aires, Argentina}

\date{\today}

\begin{abstract}
    
Hawking radiation is a fundamental result of quantum field theory in curved spacetime, yet its direct observation remains beyond current experimental capabilities. Circuit quantum electrodynamics provides a practical platform for realizing analogue systems where Hawking-like  radiation may be studied under controlled laboratory conditions. In this work, we analyze two superconducting-circuit analogues of Schwarzschild black holes: a tunable dc-SQUID transmission line and a SNAIL-based transmission line supporting solitonic solutions of the KdV equation. We investigate the conditions under which these architectures can generate an observable Hawking temperature and study the impact of dissipation and thermal noise using an open quantum systems approach. To assess the observability of the Hawking signal, we propose complementing particle number measurements with estimates of the Hilbert-Schmidt distance to the thermal bath. Our analysis establishes practical detectability thresholds and shows that  Hawking temperatures above approximately 73 mK remain distinguishable under realistic experimental conditions. While the tunable transmission line architecture can reach temperatures of about 113 mK and therefore appears more viable, the solitonic model requires further optimization and more demanding experimental conditions.

\end{abstract}

\maketitle

\section{Introduction}\label{sec:intro}

The primary motivation behind the proposal of black hole analogues is to investigate one of the most remarkable predictions of quantum field theory in curved spacetime: Hawking radiation. According to Hawking's theory, black holes are not entirely black but emit particles with a thermal spectrum, with a temperature inversely proportional to their mass \cite{Hawking1975}. As a consequence, astrophysical black holes are expected to radiate at extremely low temperatures, say a few nanokelvins, which is typically many orders of magnitude below that of the cosmic microwave background. The direct observation of Hawking radiation therefore remains beyond current experimental capabilities.

This limitation motivated the search for laboratory systems capable of reproducing some of the essential features of black hole physics. In a seminal work, Unruh \cite{unruh} showed that sound waves propagating in a fluid undergoing a subsonic-supersonic transition can be described by an effective curved spacetime endowed with a Painlevé-Gullstrand-Lemaître (PGL) metric \cite{Painleve, Gullstrand, Lemaitre}

\begin{equation}\label{eq:intervaloPGL}
    ds^2 = - \left( c_s^2 - v^2(\vec x) \right)dt^2 - 2 dtd\vec x\cdot \vec v(\vec x) + d \vec x ^2,
\end{equation}

\noindent where $c_s$ is the speed of sound and $\vec v (\vec x)$ the flow velocity. A horizon forms where $c_s^2 = v^2(\vec x)$, since acoustic perturbations can no longer move opposite to the flow. The dynamics of this perturbation is equivalent to that of a massless scalar field propagating in curved spacetime near a black hole, thereby providing a laboratory platform for the study of horizon-induced phenomena such as Hawking radiation. This construction can be generalized to the complementary situation, in which the flow velocity remains constant $v$  and the speed of sound spatially varies  $c_s(\vec x)$ \cite{javedphd}. 

Following Unruh's proposal, a plethora of analogue systems have sprung up in numerous fields, such as ion rings \cite{cirac_iones, turiaci}, Bose-Einstein condensates (BECs) \cite{Garay2000, steinhauer2016}, Fermi liquids \cite{Giovanazzi2005}, waveguides \cite{blencowe, 6nation, BH_discreto}, optical systems \cite{Unruh2003, Felipe-Elizarraras2026, medicion_HR}, polariton fluids \cite{Falque}, among others.
Their common objective is to produce experimentally accessible signatures of Hakwing radiation despite the presence of noise and other experimental imperfections. Several of these platforms have achieved remarkable experimental success.

In particular, spontaneous analogue Hawking radiation has been observed in BECs \cite{steinhauer2016} where correlations between particles across the analogue horizon were measured. Such correlations are essential for establishing the quantum nature of the effect. A discrete analogue was also measured, with a fermionic lattice-model-type realization of an analogue black hole using a chain of 10 superconducting transmon qubits \cite{BH_discreto}. Most recently, signatures of analogue Hawking radiation have also been observed in a fiber optics platform \cite{medicion_HR}, where light pulses are used to establish artificial event horizons. Also in fiber optics, an experimental verification of analogue Hawking radiation stimulated by a single-particle state, producing a verifiably single-photon Hawking output, was also realized \cite{Felipe-Elizarraras2026}. 
 
With the advent of highly controllable quantum technologies, new platforms have emerged for the study of analogue gravity phenomena. In this context, circuit quantum electrodynamics (cQED) has established itself as a versatile experimental platform, enabling the implementation of a wide range of interactions in a highly tunable setting \cite{nation2012}, while also providing powerful tools for the quantum simulation of multiple physical systems \cite{wilhelm, wilhelm_particle}. In particular, the experimental demonstration of the Dynamical Casimir Effect \cite{casimir} has  paved the way for the exploration of quantum phenomena associated with acceleration and effective horizons, such as Unruh and Hawking effects \cite{unruh-schutzhold}. Motivated by these developments \cite{turiaci, turiaci2}, we revisit transmission-line analogues of Hawking radiation within a cQED framework and investigate the role of dissipation using an open-system description.

Particularly, we focus on those analogues consisting of transmission lines where a classical magnetic flux gives rise to an effective geometry \cite{nation,katayama}. In these systems, perturbations of the magnetic flux propagate in the presence of   an effective horizon, and obey dynamics analogous to that of sound waves in Unruh's hydroacustic analogue.

Since we are interested in the quantum nature of the Hawking effect, we investigate the impact of environmental coupling on these transmission lines analogues and the resulting decoherence of the quantum field. In the original proposals mentioned above, environmental effects were not taken into account, despite the fact that they may play an important role in a realistic experimental implementation.
The presence of an environment can obscure the quantum nature of a physical system. Indeed, all real-world quantum systems interact with their surroundings  to some extent, leading to decoherence and dissipation. Even weak environmental couplings eventually induce non-unitary dynamics and  degrade quantum coherence. In practice, such effects may arise from losses and imperfections inherent to the transmission line implementation itself, as well as from the finite efficiency and noise associated with the detection process. Understanding and controlling these effects is therefore essential for assessing the feasibility of observing Hawking radiation as a genuinely quantum phenomenon in analogue-gravity  models.

In this work, we investigate whether analogue Hawking radiation in two superconducting circuit implementations can survive dissipation and thermal noise long enough to remain experimentally observable. To this end, we revisit the transmission line analogues proposed in Refs. \cite{nation,katayama} and analyze their behavior within an open quantum system framework. In particular, we study the impact of dissipation on the Hawking spectrum and establish quantitative criteria for its detectability based on both particle-number measurements and the Hilbert-Schmidt distance to the thermal bath.

The remainder of  this article is organized as follows. In Sec. \ref{sec:campo_analogo} we briefly review the emergence of a thermal spectrum from quantum fields propagating in an effective spacetime with a horizon, focusing on the relation between the \emph{in} and \emph{out} modes. In  Sec. \ref{sec:implementacion}, we discuss the analogue realization of this mechanism in the two transmission line implementations \cite{nation, katayama}. In Sec. \ref{sec:disipacion}, we extend these models to include dissipative effects and analyze their impact on the analogue Hawking radiation. Finally, we present our conclusions in Sec. \ref{sec:conclusiones}.

\section{Field treatment in analogue systems}\label{sec:campo_analogo}

In analogue-gravity systems, the dynamics of the relevant excitations can be reformulated in terms of an effective spacetime description. In particular, the field can be mapped onto a Klein-Gordon equation written in an effective curved metric, so that

\begin{equation}\label{eq:KG}
    \frac{1}{\sqrt{-g}}\partial_\mu \left( \sqrt{-g} ~ g^{\mu\nu} \partial_\nu \phi \right) = -m^2 \phi,
\end{equation}

\noindent where $g_{\mu\nu}$ is of the PGL kind as shown in Eq. \eqref{eq:intervaloPGL}. In general relativity, the PGL metric provides an intuitive fluid-like interpretation of black-hole spacetimes. In this picture, the space flows radially inward with a position-dependent velocity: outside the horizon the flow speed is smaller than the speed of light, while at the horizon both coincide. Inside the horizon, the inward flow exceeds the speed of light, preventing outward-directed light rays from escaping to infinity. A similar situation arises in the systems we shall consider as black hole transmission line analogues: by appropriately choosing the circuit parameters, one can create an effective horizon that causally separates two regions of the transmission line.

The presence of this event horizon has remarkable consequences on field quantization. In lightcone coordinates, the ingoing modes $v$, which move in the same direction as the fluid, will be naturally well defined; however, we must include two distinct sets of outgoing modes $u$: one for each of the two regions delimited by the horizon. This is necessary since for any outgoing packet outside of the horizon, when traced back in time, can never have been in the region on the opposite side. Thus, an additional set of outgoing modes are required inside the horizon to describe all of the field's degrees of freedom. \cite{43javed}. If we consider the asymptotic past and future of the field, which we will call \textit{in} and \textit{out} regions respectively, it is clear that the $v$ modes are the same in both regions since they are well defined at the horizon. For the $u$ modes this is not the case, which means that for each region we have the decomposition

\begin{equation}
    \begin{split}
    \hat{\phi} & = \int_{0}^{\infty} d\omega \,\Bigl\{\hat{a}^{u,\mathrm{in}}_{\omega,R}\phi^{u,\mathrm{in}}_{\omega,R} + \hat{a}^{u,\mathrm{in}}_{-\omega,L}\phi^{u,\mathrm{in}}_{-\omega,L} + \hat{a}^{v}_{\omega}\phi^{v}_{\omega} + \mathrm{h.c.}\Bigr\} \\
    & = \int_{0}^{\infty} d\omega \,\Bigl\{\hat{a}^{u,\mathrm{out}}_{\omega,R}\phi^{u,\mathrm{out}}_{\omega,R} + \hat{a}^{u,\mathrm{out}}_{-\omega,L}\phi^{u,\mathrm{out}}_{-\omega,L} + \hat{a}^{v}_{\omega}\phi^{v}_{\omega} + \mathrm{h.c.}\Bigr\}.
    \end{split}
\end{equation}

\vspace{-10pt}

Requiring that the $u$ modes connect to each other analytically through the horizon \cite{43javed} it can be readily shown that the \textit{in} and \textit{out} ladder operators must be related to each other by \cite{7de43javed, 15de43javed, 22de43javed}

\begin{equation}
    \begin{split}
        \hat{a}^{u,\mathrm{out}}_{\omega,R} & = \frac{1}{\sqrt{2\sinh\!\left(\pi\omega/c_h'\right)}}\left(e^{\frac{\pi\omega}{2c_h'}}\hat{a}^{u,\mathrm{in}}_{\omega,R} + e^{-\frac{\pi\omega}{2c_h'}}\hat{a}^{u,\mathrm{in}\dagger}_{-\omega,L}\right),\\
        \hat{a}^{u,\mathrm{out}}_{-\omega,L} & = \frac{1}{\sqrt{2\sinh\!\left(\pi\omega/c_h'\right)}}\left(e^{\frac{\pi\omega}{2c_h'}}\hat{a}^{u,\mathrm{in}}_{-\omega,L} + e^{-\frac{\pi\omega}{2c_h'}}\hat{a}^{u,\mathrm{in}\dagger}_{\omega,R}\right);
    \end{split}
\end{equation}

\noindent where $c_h'$ is the first derivative of $c_s(\vec x)$ evaluated at the horizon. This allows us to compute the number of particles outgoing from the horizon in the \textit{out} state assuming the \textit{in} state to be vacuum, through 

\begin{equation}\label{eq:n_out}
    \begin{split}
        \langle n^{u,\mathrm{out}}_{\omega,R}\rangle & = \langle 0_{\mathrm{in}}|\hat{a}^{u,\mathrm{out}\dagger}_{\omega,R}\hat{a}^{u,\mathrm{out}}_{\omega,R}|0_{\mathrm{in}}\rangle\\
        & = \frac{e^{-\pi\omega/c_h'}}{2\sinh\!\left(\pi\omega/c_h'\right)}\langle0_{\mathrm{in}}|\hat{a}^{u,\mathrm{in}}_{\omega,L}\hat{a}^{u,\mathrm{in}\dagger}_{\omega,L}|0_{\mathrm{in}}\rangle \\
        & = \frac{1}{e^{2\pi\omega/c_h'}-1}.
    \end{split}
\end{equation}

We thus have a spectrum of emitted quasiparticles which corresponds to a thermal Bose–Einstein distribution. The corresponding Hawking temperature can be inferred from Eq. \eqref{eq:n_out}. Assuming that the field propagates in the $x$  direction, it takes the more general form \cite{6nation}

\begin{equation}\label{eq:TH_nation}
    T_H = \frac{\hbar}{2\pi k_B} \left| \frac{\partial c_s(x)}{\partial x} \right|_{c_s^{2} = v^{2}}.
\end{equation}

This result implies that  any continuous system whose field can, in some parameter regime, be described by Eq. \eqref{eq:KG} with a PGL metric and a subsonic-supersonic transition will exhibit analogue Hawking radiation with a temperature given by Eq. \eqref{eq:TH_nation}. These systems therefore offer a controlled setting in which to investigate the quantum-field-theoretic mechanisms underlying Hawking radiation.

The transmission-line systems considered in this work constitute particular realizations of this general framework. In the next section, we discuss how the effective metric and the corresponding horizon emerge in transmission line Black Hole analogues.

\section{Transmission line Black Hole analogues in superconducting circuits}\label{sec:implementacion}

In this section we introduce the two superconducting- circuit implementation that will be considered throughout this work. Both proposals realize effective one dimensional analogues of Schwarzschild black holes. The first, proposed by P. D. Nation et. al. \cite{nation} and R. Schützhold and W. G. Unruh \cite{unruh-schutzhold}, relies on the modulation of the propagation velocity of the relevant field through an external input. We then discuss an alternative approach by H. Katayama et. al. \cite{katayama} in which the efective analogue geometry emerges from solitonic solutions of the Korteweg-de Vries (KdV) and modified KdV (mKdV) equations.

\subsection{Tunable transmission line analogue}\label{sec:nation}

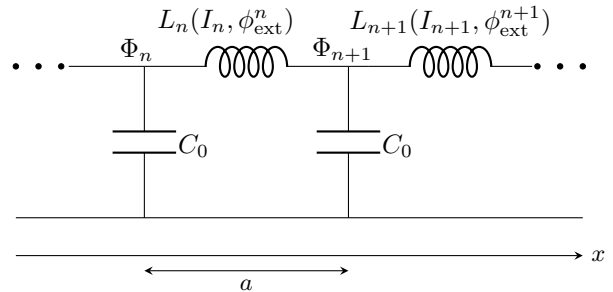
\begin{figure} [h!]
    \centering
    \begin{circuitikz}
        \def\a{2.7} 
        \draw[fill=black] (-1.1 + \a,2) circle (1pt);
        \draw[fill=black] (-1.7 + \a,2) circle (1pt);
        \draw[fill=black] (-1.4 + \a,2) circle (1pt);
        \draw (-1 + \a,2) -- (0 + \a,2);
        \draw (\a,2) to[L, style={font=\normalsize}, bipoles/length=1.75cm] (2*\a,2);
        \node[style={font=\normalsize}] at (\a + 0.4*\a,2.6) {$L_n(I_{n}, \phi^{n}_{\text{ext}})$};
        \node[style={font=\normalsize}] at (\a - 0.1,2.22) {$\Phi_n$};
        \draw (\a,0) to[capacitor] (\a,2);
        \node[style={font=\normalsize}] at (\a + 0.65,0.95) {$C_{0}$};
        \draw[-] (2*\a,2) -- (2.1*\a,2);
        \draw (2.1*\a,2) to[L, style={font=\normalsize}, bipoles/length=1.75cm] (2.9*\a,2);
        \node[style={font=\normalsize}] at (2*\a + 0.5*\a,2.6) {$L_{n+1}(I_{n+1}, \phi^{n+1}_{\text{ext}})$};
        \node[style={font=\normalsize}] at (2*\a - 0.08,2.22) {$\Phi_{n+1}$};
        \draw (2*\a,0) to[capacitor] (2*\a,2);
        \node[style={font=\normalsize}] at (2*\a + 0.65,0.95) {$C_{0}$};
        \foreach \x in {1.1, 1.4, 0.8} {
          \draw[fill=black] (3*\a+\x-1,2) circle (1pt);
        }
        \draw[-] (-1.7 + \a,0) -- (3*\a+1.7-1.3,0) node {};
        \draw[-stealth] (-1.7 + \a,-0.5) -- (3*\a+1.7-1.3,-0.5) node[right] {$x$};
        \draw[stealth-stealth] (\a,-0.7) -- (2*\a,-0.7) node[midway,below] {$a$};
    \end{circuitikz}
    \caption{Lumped circuit model of a transmission line composed of continuous horizontal dc-SQUIDs in a low-current and low-frequency regime, as proposed by \cite{nation}. In this regime, each SQUID acts as an inductor that can be space and time tuned through an external flux and current bias \cite{unruh-schutzhold}.}
    \label{fig:circuitoNation}
\end{figure}

We begin with the first analogue, namely the tunable coplanar waveguide shown in Fig. \ref{fig:circuitoNation}, realized as an array of  $N \gg 1$ identical dc-SQUIDs placed horizontally next to one another \cite{nation, unruh-schutzhold}. As is well known, the current through a SQUID with negligible self-inductance is given by

\begin{equation}\label{eq:I_SQUID}
    I = 2 C_J \frac{\Phi_0}{2\pi} \frac{d^2\phi}{dt^2} + 2 I_c \cos \left( \frac{\pi \Phi_{\text{ext}}}{\Phi_0} \right) \sin (\phi),
\end{equation}

\noindent where $I_c$ and $C_J$ are the junction critical current and capacitance, respectively; $\Phi_0 = h/2e$ is the flux quantum and $\phi = \Phi/\Phi_0$ is the phase variable. Here, $\Phi$ is the electromagnetic flux across the transmission line.

Assuming that the  current through the junction remains much smaller than the critical current and that the characteristic frequencies of the electromagnetic flux are well below the plasma frequency $\omega_p^2 = 2\pi I_c/C_J\Phi_0$, the dc-SQUID behaves as a flux-dependent inductance given by

\begin{equation}\label{eq:LJ}
    L = \frac{\Phi_0}{2\pi I}\phi = \frac{\Phi_0}{2\pi I_c^s} \frac{\arcsin(I/I_c^s)}{I/I_c^s},
\end{equation}

\noindent where we have defined the effective critical current $I_c^s = 2I_c \cos \left( \pi \Phi_{\text{ext}}/\Phi_0 \right)$. Equations \eqref{eq:I_SQUID} and \eqref{eq:LJ} show that the dc-SQUID  behaves as an effective Josephson junction, whose critical current, and consequently its inductance, can be tuned through the external magnetic  flux. 

Applying Kirchhoff's laws to the circuit in Fig. \ref{fig:circuitoNation}, one finds that each node satisfies $I_{n-1} = I_n + C_0\dot V_n$. Using that $V = \dot \Phi$ and Eq. \eqref{eq:LJ}, this yields

\begin{equation}
    \frac{\Phi_n - \Phi_{n-1}}{L_{n-1}} = \frac{\Phi_{n+1} - \Phi_{n}}{L_{n}} - C_0\frac{d^2\Phi_n}{dt^2}.
\end{equation}

To complement previous work, we propose in this work  to obtain the same circuit dynamics from a Lagrangian formulation written in terms of the node fluxes, which provides a convenient starting point for quantization. In the long-wavelength regime, where all relevant wavelengths satisfy $\lambda \gg a$ (where $a$ is the physical length of a single SQUID element, i.e., the array unit cell size), the discrete SQUID array can be  approximated as a continuous transmission line. The flux variable then becomes a continuous electromagnetic flux field $\Phi(x,t)$ governed by

\begin{equation}\label{eq:LagrangianoNation}
    \mathcal{L} = \frac{C_0}{a}\int dx \left\{ \frac{\left(\partial_t\Phi\right)^2}{2} - \frac{a^2}{2C_0L(x,t)} \left(\partial_x\Phi\right)^2 \right\},
\end{equation}

\noindent from which we identify $c^2(x,t) = a^2/C_0L(x,t)$ as the tunable propagation speed of the field $\Phi$. 

Since the external flux bias $\Phi_{\text{ext}}$ can be controlled by applying a magnetic field to the transmission line, the propagation speed may be engineered to acquire a space and time dependent profile of the form $c(x,t) = c(x-u_0t)$, corresponding to a pulse traveling at speed $u_0$. We can then change coordinates to ones comoving with this pulse, that is, $\xi = x - u_0 t$. In the comoving frame, after relabeling $\xi \rightarrow x$, the Lagrangian becomes

\begin{equation}\label{eq:LagrangianoNationComovil}
    \mathcal{L} = \frac{C_0}{a}\int dx \left\{ \frac{1}{2}\left(\partial_t\Phi - u_0\partial_x\Phi\right)^2 - \frac{1}{2}c^2(x) \left(\partial_x\Phi\right)^2 \right\}.
\end{equation}

In 1+1 dimensions, however, the scalar field $\Phi$ is conformally invariant. As a consequence, the effective geometry cannot be uniquely identified from the Lagrangian alone, since different metrics related by a conformal transformation lead to the same field equation. Nevertheless, the physical waveguide is not strictly one-dimensional. Although the transverse dimension satisfies $\lambda \gg a_y \gg a$ and therefore does not contribute to the low-energy dynamic \cite{6nation}, its presence allows the identification of a well-defined effective spacetime geometry. Taking this extra dimension into account, the field equation can be written in the form of Eq. \eqref{eq:KG} with the effective metric

\begin{equation}\label{eq:g_munu}
    g^{\mu\nu}_{\text{eff}} = \frac{1}{c^4(x)}
    \begin{pmatrix}
        1 & -u_0 & 0 \\
        -u_0 & u_0^2 - c^2(x) & 0 \\
        0 & 0 & -c^2(x)
    \end{pmatrix}.
\end{equation}

Since $a_y \gg a$, the transverse wavenumber $k_y$ is discrete. Furthermore, the condition $\lambda \gg a_y$ implies that the spacing between transverse modes is sufficiently large that only the lowest mode is appreciably populated. Under these conditions, the field dynamics is effectively restricted to the transverse ground mode, so that the additional dimension does not contribute to the low-energy dynamics.

The effective metric of Eq. \eqref{eq:g_munu} is of the Painlevé-Gullstrand-Lemaiîre kind and leads to a line element analogous to  Eq. \eqref{eq:intervaloPGL}. Here, we may choose an external flux such that $c(x)<u_0$ for $x<x_h$ and $c(x)>u_0$ for $x>x_h$, effectively creating an horizon at $x=x_h$. By choosing the profile $c(x)$ to cross $u_0$ only once with positive slope at the crossing, we can ensure that there is only a black hole and not a white one.

As previously discussed, the presence of the horizon at $x=x_h$ leads to the emission of a thermal spectrum of analogue Hawking radiation propagating along the transmission line. In addition, the condition $u_0 < c_0 = c(\Phi_{\text{ext}}=0)$ must be met, otherwise Hawking radiation would be unable to propagate to the asymptotic \textit{out} region. 

Since the above result is obtained in the frame comoving with the pulse, the temperature measured in the rest frame will be Doppler shifted by a factor of $\sqrt{(c_0-u_0)/(c_0+u_0)}$. To the lowest order in $I/I_c^s$, it can be shown that the propagation speed of the field in the transmission line can be expressed as

\begin{equation}
    c^2(x) \approx \frac{4\pi a^2 I_c}{C_0 \Phi_0} \cos \left( \frac{\pi \Phi_{\text{ext}}}{\Phi_0} \right) = c_0^2 \cos \left( \frac{\pi \Phi_{\text{ext}}}{\Phi_0} \right).
\end{equation}

\begin{figure} [h!]
    \centering
    \includegraphics[width=\linewidth]{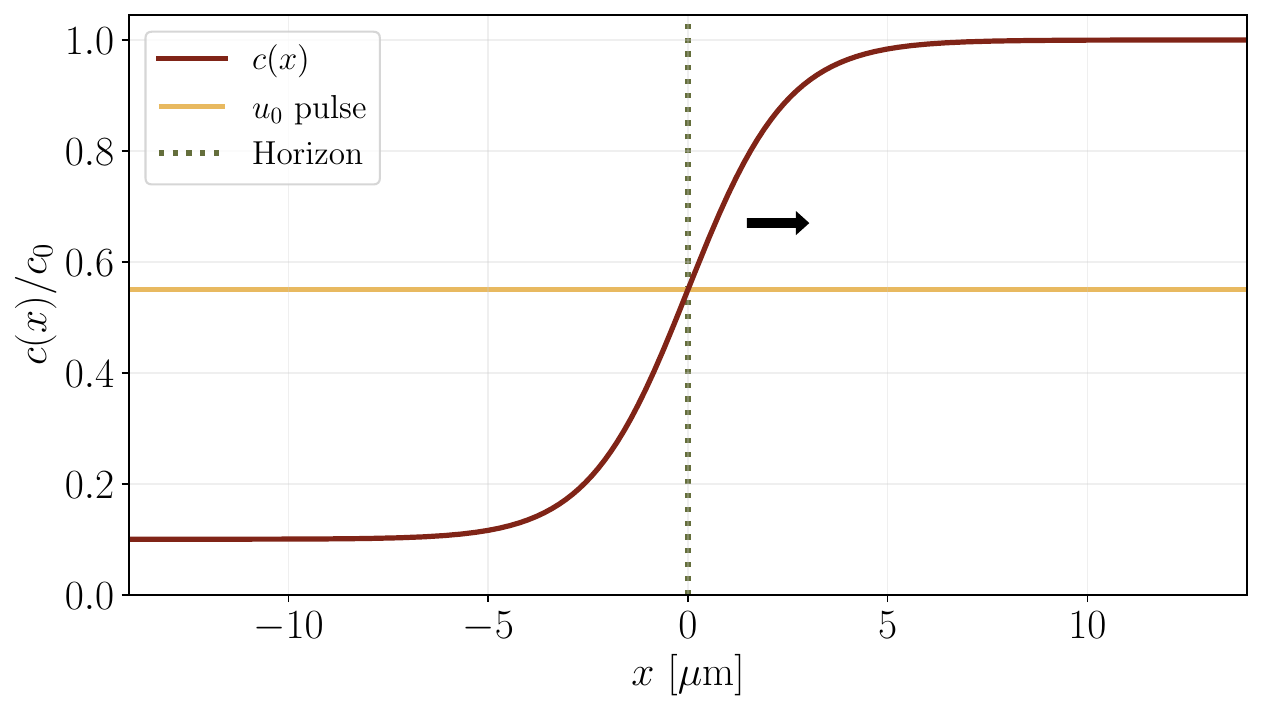}
    \caption{Propagation velocity profile $c(x)$ at time $t=0$ in units of $c_0$ across the transmission line. The horizon occurs where $c(x) = u_0$. The arrow indicates the direction in which the pulse propagates in the laboratory frame.}
    \label{fig:perfil_c(x)}
\end{figure}

In Fig. \ref{fig:perfil_c(x)} we show a representative spatial profile for the propagation speed of the electromagnetic flux-field excitations of the form

\begin{equation}\label{eq:perfil}
    c(x) = v\tanh\left( x/\lambda_c \right) + u_0,
\end{equation}

\noindent where $u_0+v$ and $u_0-v$ are the asymptotic values of $c(x)$ as $x \to \infty$ and $x \to -\infty$, respectively. The horizon is located at $x=0$, where the gradient of the profile is maximal. The parameter $v$ is constrained by $v\leq c_0-u_0$ to ensure that the profile admits a single horizon and that $c(x)$ interpolates between well-defined asymptotic values. Finally, $\lambda_c$ is the characteristic wavelength of the external flux, and we assume $\lambda_c \gg a$ in order for the continuum approximation to remain valid.

When computing the Hawking temperature for the profile in Eq. \eqref{eq:perfil} using Eq. \eqref{eq:TH_nation}, we obtain

\begin{equation}\label{eq:TH_perfil}
    T_H = \frac{\hbar v}{2\pi k_B\lambda_c},
\end{equation}

\noindent where $v/\lambda_c$ imposes a frequency scale \cite{nation} in the following sense: the thermal spectrum produced by the radiated particles is

\begin{equation}\label{eq:thermal_spectrum}
    n(\omega) = \frac{1}{e^{\hbar\omega/k_BT_H}-1},
\end{equation}

\noindent which means that all relevant frequencies are of order $k_BT_H/\hbar$ or less. As we can see from Eq. \eqref{eq:TH_perfil}, $k_BT_H/\hbar \propto v/\lambda_c$ which makes this ratio precisely the relevant frequency scale. Since we are working with frequencies well below the plasma frequency of the circuit, which takes typical values of $10^{12}$ Hz, this profile can reach a maximum theoretical Hawking temperature of around 120 mK for $v/\lambda_c = 10^{11}$ Hz. Since cryostats available today can be cooled to temperatures as low as 10 mK, the predicted Hawking temperature lies well above the thermal background, suggesting that the emitted radiation could be experimentally accessible.

Nevertheless, care must be taken when choosing the circuit parameters. In order to maximize the Hawking temperature one would like to make the ratio $v/\lambda_c$ as large as possible. However, the propagation speed profile must remain positive and bounded by the maximum attainable value $c_0$, which means that $v\leq u_0$ and $v\leq c_0-u_0$ must be satisfied. In addition, the continuum approximation requires $\lambda_c$ considerably larger than $a$. Under these constraints, an optimal choice is $v=u_0=c_0/2$. Assuming $\lambda_c=10a$, and using $c_0/a \approx 2.46\cdot10^{12}~ \rm Hz$ one finds $v/\lambda_c=c_0/20a \approx 1.23 \cdot 10^{11}\rm Hz$ corresponding to a Hawking ideal temperature of approximately $120~\rm mK$.

For a representative set of circuit parameters:  $a=0.25$ $\mu$m, $I_c = 2$ $\mu$A, $C_0 = 2.87$ fF, $\lambda_c = 10a$, $u_0 = 0.55c_0$ and $v=0.45c_0$ \cite{nation}, the resulting thermal spectrum produces a Hawking temperature of 113 mK, close to this theoretical limit.

\subsection{Solitonic transmission line analogue}

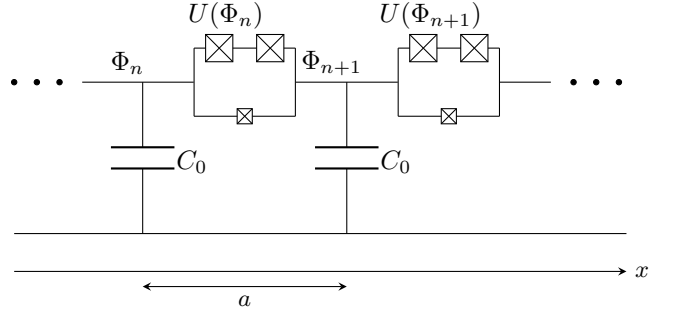
\begin{figure}[h]
    \centering
    \begin{circuitikz}
        \def\a{2.7} 
        \newcommand{\snail}[2]{%
            \draw (#1-0.67, #2-0.45) -- (#1-0.67, #2+0.45);
            \draw (#1+0.67, #2-0.45) -- (#1+0.67, #2+0.45);
            \draw (#1-0.67, #2+0.45) -- (#1*0.75-0.18, #2+0.45);
            \draw (#1*0.75+0.18, #2+0.45) -- (#1*1.25-0.18, #2+0.45);
            \draw (#1*1.25+0.18, #2+0.45) -- (#1+0.67, #2+0.45);
            \draw (#1-0.67, #2-0.45) -- (#1-0.1, #2-0.45);
            \draw (#1+0.1, #2-0.45) -- (#1+0.67, #2-0.45);
            \draw (#1*1.25-0.18, #2+0.45-0.18) -- (#1*1.25+0.18, #2+0.45+0.18);
            \draw (#1*1.25-0.18, #2+0.45+0.18) -- (#1*1.25+0.18, #2+0.45-0.18);
            \draw (#1*1.25-0.18, #2+0.45-0.18) rectangle (#1*1.25+0.18, #2+0.45+0.18);
            \draw (#1*0.75-0.18, #2+0.45-0.18) -- (#1*0.75+0.18, #2+0.45+0.18);
            \draw (#1*0.75-0.18, #2+0.45+0.18) -- (#1*0.75+0.18, #2+0.45-0.18);
            \draw (#1*0.75-0.18, #2+0.45-0.18) rectangle (#1*0.75+0.18, #2+0.45+0.18);
            \draw (#1-0.1, #2-0.45+0.1) -- (#1+0.1, #2-0.45-0.1);
            \draw (#1-0.1, #2-0.45-0.1) -- (#1+0.1, #2-0.45+0.1);
            \draw (#1-0.1, #2-0.45+0.1) rectangle (#1+0.1, #2-0.45-0.1);
        }
        \draw[fill=black] (-1.1 + \a, 2) circle (1pt);
        \draw[fill=black] (-1.7 + \a, 2) circle (1pt);
        \draw[fill=black] (-1.4 + \a, 2) circle (1pt);
        \snail{\a + 0.5*\a}{ 2 }
        \draw (1.75*\a, 2) -- (2.25*\a , 2);
        \draw (\a - 0.8, 2) -- (1.25*\a, 2);
        \node[font=\normalsize] at (1.5*\a-0.25, 2.88)
            {$U(\Phi_{n})$};
        \node[font=\normalsize] at (\a - 0.2, 2.25) {$\Phi_n$};
        \draw (\a, 0) to[capacitor] (\a, 2);
        \node[font=\normalsize] at (\a + 0.65, 0.95) {$C_{0}$};
        \def\b{2*\a}   
        \snail{2*\a + 0.5*\a}{ 2 }
        \draw (2.75*\a, 2) -- (3*\a, 2);
        \node[font=\normalsize] at (2.5*\a - 0.24, 2.88)
            {$U(\Phi_{n+1})$};
        \node[font=\normalsize] at (2*\a - 0.2, 2.25) {$\Phi_{n+1}$};
        \draw (2*\a, 0) to[capacitor] (2*\a, 2);
        \node[font=\normalsize] at (2*\a + 0.65, 0.95) {$C_{0}$};
        \foreach \x in {0.0, 0.3, 0.6} {
            \draw[fill=black] (3*\a + \x + 0.3, 2) circle (1pt);
        }
        \draw (-1.7 + \a, 0) -- (3*\a + 1, 0);
        \draw[-stealth] (-1.7 + \a, -0.5) -- (3*\a + 1, -0.5) node[right] {$x$};
        \draw[stealth-stealth] (\a, -0.7) -- (2*\a, -0.7) node[midway, below] {$a$};
    \end{circuitikz}
    \caption{Lumped circuit model of a transmission line composed of continuous horizontal SNAILs. The junctions at the top of this element have energy $E_J$ while the one at the bottom has energy $\alpha E_J$.}
    \label{fig:SNAIL}
\end{figure}

To realize the solitonic analogue, Ref. \cite{katayama} considers a modified version of the transmission line shown in Fig. \ref{fig:circuitoNation}, where the dc-SQUIDs are replaced by Superconducting Nonlinear Asymmetric Inductive eLements (SNAILs). A SNAIL consists of a superconducting loop interrupted by three Josephson junctions: 
 one  branch contains two junctions with Josephson energy $E_J$, while the other contains a single junction with Josephson energy  $\alpha E_J$ ($\alpha < 1$), as shown in Fig. \ref{fig:SNAIL}.

This element has an energy composed of a kinetic capacitive term and a potential term given by \cite{snail}

\begin{equation}\label{eq:U_snail_exacto}
    U(\phi) = - \alpha E_J \cos(\phi) - 2 \cos\left( \frac{\phi - \phi_{\text{ext}}}{2} \right),
\end{equation}

\noindent where $\phi_{\text{ext}} = 2\pi\Phi_{\text{ext}}/\Phi_0$. When expanded around its minimum value, found numerically, this potential takes the shape

\begin{equation}\label{eq:Usnail}
    U(\phi) \approx E_J \left[ \frac{\alpha(\phi_{\text{ext}})}{2!}\tilde\phi^2 + \frac{\beta(\phi_{\text{ext}})}{3!}\tilde\phi^3 + \frac{\gamma(\phi_{\text{ext}})}{4!}\tilde\phi^4 \right].
\end{equation}

As in the tunable transmission-line analogue discussed above, the circuit dynamics can be derived by applying Kirchhoff's laws to each node of the array. The resulting equations of motion contain a capacitive contribution together with the nonlinear force obtained from the above potential. Combining the equations for neighboring nodes and taking the continuum limit, one obtains

\begin{equation}\label{eq:katayamacontinuo}
        \frac{\partial^2\phi}{\partial t^2} - a^2 r \frac{\partial^4 \phi}{\partial t^2 \partial x^2}  - v_0^2 \frac{\partial^2}{\partial x^2} \left( \phi + \frac{c_3}{2}\phi^2 + \frac{c_4}{6}\phi^3 \right) = 0,
\end{equation}

\noindent where $r=C_J/C_g$, $v_0^2 = a^2\omega_0^2 = a^2/C_gL_0$ with $L_0 = \varphi_0/I_c\alpha(\phi_{\text{ext}})$, $c_3 = \beta(\phi_{\text{ext}})/\alpha(\phi_{\text{ext}})$ and $c_4 = \gamma(\phi_{\text{ext}})/\alpha(\phi_{\text{ext}})$. 

Following \cite{katayama}, we decompose the flux field as $\phi = \bar\phi + \delta\phi$, where $\bar\phi$ is a background solution of Eq. \eqref{eq:katayamacontinuo} and $\delta\phi$ is a small fluctuation propagating on top of it. We first seek background solutions that propagate along the transmission line without changing shape, namely solitons. Such solutions arise when the competing effects of dispersion and nonlinearity are balanced. To derive the effective nonlinear evolution equation controlling this regime, we employ the reductive perturbation method \cite{30katayama,gardner_morikawa}, introducing the ``stretched'' variables through the Gardner-Morikawa transformations, given by

\begin{equation}
    \begin{split}
        \chi = \varepsilon^{1/2}(x - v_0t), &\quad
        \tau = \varepsilon^{3/2}t, \\
        \bar\phi = \varepsilon^{i}\phi^{(1)} + \varepsilon^{2i}&\phi^{(2)} + \cdot\cdot\cdot 
    \end{split}
\end{equation}

\noindent where $i$ is a rational number fixed by requiring that the leading  nonlinear and dispersive contributions scale in the same order in $\varepsilon$. By substituting the above equations into Eq. \eqref{eq:katayamacontinuo}, with $i\leq1$ we obtain a consistent ordering in $\varepsilon$ in which higher order time derivatives are suppressed and can be neglected at leading order. In addition, we set
either $c_3=0$ or $c_4=0$ in order to isolate cubic or quartic non-linearities, respectively.

At leading order in $\varepsilon$, this procedure yields either the Korteweg-de Vries (KdV) equation or one of the modified KdV equations (mKdV$\pm$), depending on the choice of $i$. These effective equations are scale invariant and admit solitonic solutions, namely localized wave packets that propagate at constant velocity without changing their shape. As an example, the KdV soliton is given by

\begin{equation}\label{eq:phi_kdv}
    \bar\phi_{\text{KdV}} = A \, \text{sech}^2\left[ \sqrt{\frac{c_3A}{12ra^2}}(x-v_st) \right],
\end{equation}

\noindent obtained by choosing $i=1$ and values of ($\alpha$, $\phi_{\text{ext}}$) such that $c_4=0$. For this soliton the velocity is $v_s = v_0(1 + c_3A/6)$, with $c_3A > 0$ \cite{katayama}. 

Having obtained the background soliton solution, we now turn to the dynamics of the fluctuation field  $\delta\phi$, whose first order dynamics is ruled by the equation

\begin{equation}\label{eq:katayamadphi}
        \frac{\partial^2\delta\phi}{\partial t^2} - a^2 r \frac{\partial^4\delta\phi}{\partial t^2 \partial x^2}  - \frac{\partial^2}{\partial x^2} \left( v^2(x,t) \ \delta\phi \right) = 0.
\end{equation}

Here we define the variable $\eta = x - v_st$ comoving with the soliton. Introducing the auxiliary field $\delta \varphi$ through $\delta\phi = a (d\delta\varphi/d\eta)$ and considering the long wavelength limit, it can be shown that $\delta\varphi$ obeys

\begin{equation}
    \left\{ \frac{\partial^2}{\partial t^2} - 2v_s\frac{\partial^2}{\partial t \partial \eta} + \frac{\partial}{\partial\eta} \left[ \left(v_s^2 - v^2(\eta)\right) \frac{\partial}{\partial\eta}  \right] \right\} \delta\varphi = 0.
\end{equation}

\noindent Under the same assumptions discussed for the tunable transmission-line analogue, this equation can be cast into the form of a Klein-Gordon equation in an effective curved spacetime with a metric analogous to Eq. \eqref{eq:g_munu}. The fluctuation field $\delta \varphi$ therefore experiences an effective horizon structure and is expected to exhibit analogue Hawking radiation with a temperature determined by Eq. \eqref{eq:TH_nation}. In this case, however, the effective propagation speed of $\delta\varphi$ is 

\begin{equation}
    v(x, t) = v_0\sqrt{1 + c_3\bar\phi + \frac{c_4}{2}\bar\phi^2}.
\end{equation}

For all three solitonic backgrounds, the condition $v(\eta) = v_s$ is satisfied at two values of $\eta$ symmetrically located about $\eta=0$. These two crossings correspond to both a black hole  and a white hole horizon. This configuration complicates the detection of the Hawking signal, since the relevant radiation is confined to the region between the black hole and white hole horizons, rather than being emitted into an asymptotic exterior region as in the configuration studied in the previous section.

\begin{figure} [h!]
    \centering
    \includegraphics[width=\linewidth]{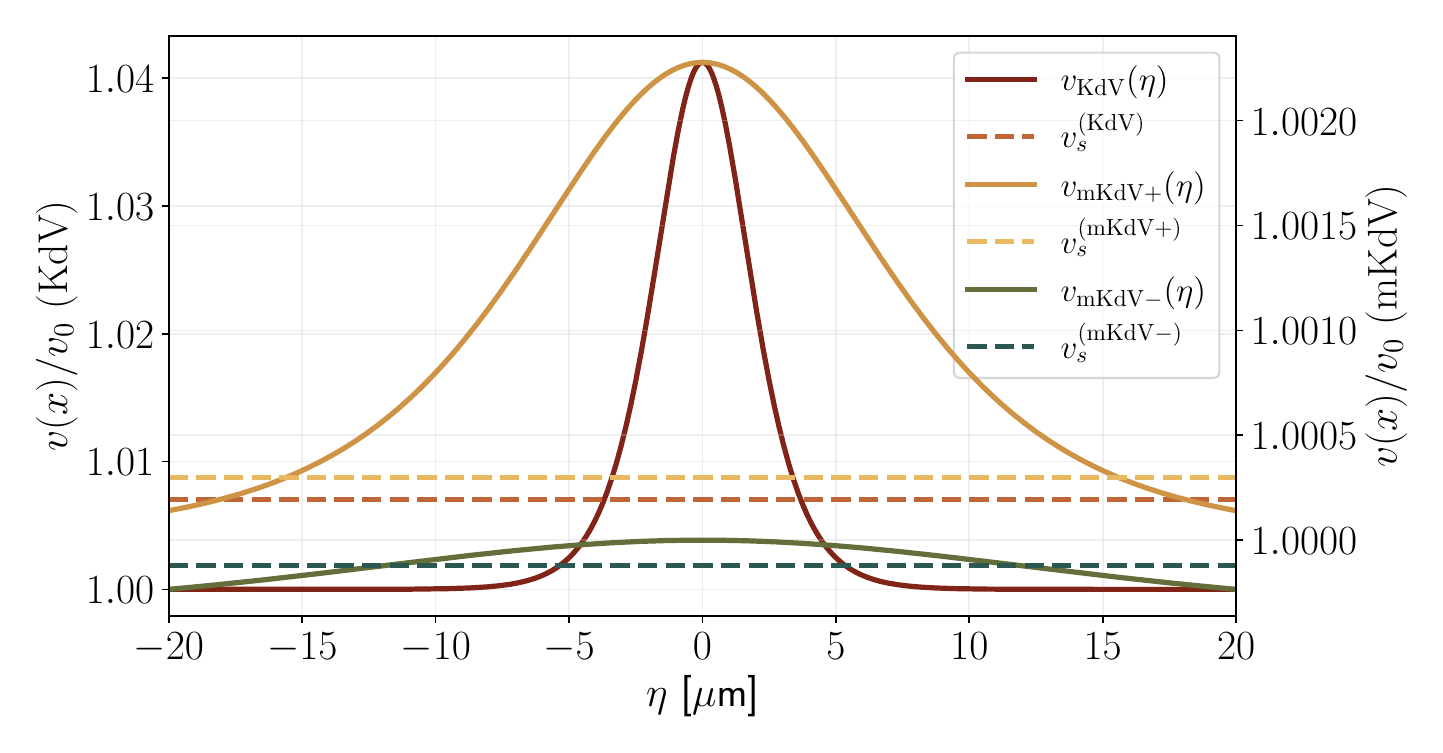}
    \caption{Field propagation velocity (solid lines) and soliton velocity (dashed lines) in the comoving frame normalized by $v_0$ for the three solitonic cases. The left $y$ axis is the scale for the KdV case, while the right one is for the mKdV cases. The left and right intersections represent the black hole and white hole horizons respectively.}
    \label{fig:v_katayama}
\end{figure}

The parameter $\alpha$ in Eq. \eqref{eq:U_snail_exacto} is fixed during fabrication of the SNAIL, and unlike $\phi_{\text{ext}}$ cannot be tuned during an experiment. We thus seek a value for $\alpha$ that allows all three solitonic analogues (KdV, mKdV+ and mKdV-) to attain large Hawking temperatures through the appropriate choices of $\phi_{\text{ext}}$. We find that  $\alpha = 0.4026$ provides a suitable compromise: by varying $\phi_{\text{ext}}$, one can realize that  $c_3 = 0.84$ and $c_4=0$ (KdV), $c_4 = 2.85$ with $c_3=0$ (mKdV+), and $c_4 = -0.58$ with $c_3=0$ (mKdV-). Fig. \ref{fig:v_katayama} shows the corresponding propagation speed profiles together with the associated soliton velocities. These parameter choices lead to higher Hawking temperatures than those reported  in \cite{katayama}.

It is clear from Fig. \ref{fig:v_katayama} that the KdV solution yields the highest Hawking temperature. This conclusion can also be reached both analytically and numerically: the Hawking temperature associated with the mKdV solitons is an order of $A$ smaller than the KdV case. For the representative parameters $a = 2$ $\mu$m, $|A|=0.05$, $I_c = 2$ $\mu$A and $C_g = C_J = 1$ fF, which are the ones used for Fig. \ref{fig:v_katayama}, we obtain $T_H^{\text{KdV}} = 1.2$ mK 
whereas $T_H^{\text{mKdV}\pm} = 0.01$ mK. Consequently, one may instead choose to optimize the transmission line exclusively for the KdV analogue. In that case, selecting $\alpha = 0.66$ allows one to reach  $c_3 = 1.74$, yielding a Hawking temperature of $T_H^{\text{KdV}} = 1.9$ mK. 

For completeness, we mention that another solitonic analogue was proposed by \cite{Tian}, where the magnetic flux obeys the Sine-Gordon equation. In this case, the soliton solutions induce an effective constant negative curvature geometry corresponding to a two-dimensional black hole of Jackiw-Teitelboim dilaton gravity. As in the previous analogues, the presence of an effective horizon leads to Hawking radiation, with a temperature proportional to the soliton velocity. For realistic circuit parameters, Hawking temperatures up to 150 mK have been predicted.

\section{Dissipative Schwarzschild analogues in a transmission line}\label{sec:disipacion}

A complete description of the quantum phenomena we have described so far requires taking into account the influence of the environment. Besides inducing decoherence and dissipation, environmental degrees of freedom also model the unavoidable coupling associated with control and measurement procedures. Since the observation of analogue Hawking radiation necessarily relies on external probing of the circuit, neglecting these interactions would be inconsistent. We thus investigate how environmental coupling affects the transmission line analogues introduced in the previous sections.

\subsection{Infinite mode field treatment}

Following the standard  cQED approach, we model the interaction between the transmission lines proposed in this paper and the environment by means of the Lindblad equation  \cite{petruccione}

\begin{equation}\label{eq:lindblad}
    \dot{ \rho}(t) = - \frac{i}{\hbar} [ H_s, \rho(t)] + \sum_{\alpha} \mathcal{D}[ L_\alpha]  \rho(t),
\end{equation}

\noindent where $\rho(t)$ is the reduced density matrix of our system, $H_s$ is the system Hamiltonian, $L_\alpha$ are the so-called Lindblad operators and $\mathcal{D}$ is the dissipator superoperator

\begin{equation}
    \mathcal{D}[ L_\alpha] \,  \bullet =  L_{\alpha} \,  \bullet \,  L_\alpha^\dagger - \frac{1}{2} \{  L_\alpha^\dagger  L_\alpha, \,  \bullet \, \}.
\end{equation}

Even though this derivation has been thoroughly studied, it is relevant in our case to remember one of the hypotheses required to reach Eq. \eqref{eq:lindblad}: the \textit{Born-Markov approximation}, or weak coupling approximation. It is assumed that the interaction between the system and the environment is weak enough so that correlations remain sufficiently weak that the total density operator can be approximated as a product state throughout the evolution. This also means that the bath evolves separately from the system of interest and, moreover, that any excitations that this reservoir gains from its interaction with the system will decay in a time scale fast enough to consider the state of the environment to be constant at all times. 

Using this equation is relatively straightforward in the case of finite-dimensional systems such as qubits. The transmission line analogues studied here, however, involve the quantization of an infinite mode field in an effective curved spacetime (Eq. \eqref{eq:g_munu}). Quantization in this regime is not unique, since the field can be expanded in terms of multiple nonequivalent positive frequency solutions. This applies equally to creation and annihilation operators, which means that fundamental notions such as particles and the vacuum state are no longer uniquely defined. Consequently, the associated creation and annihilation operators cannot always be interpreted as describing the loss or gain of physical excitations. This complicates the identification of meaningful Lindblad operators. Only in special cases, such as conformally flat or asymptotically Minkowski effective spacetimes, does a preferred particle interpretation emerge.

In the asymptotic \textit{out} region, in which our effective space-time is flat, it is reasonable to study dissipation through Lindblad's equation. Unfortunately, this is not enough to guarantee that this equation describes what we wish to model. Our field has an infinite (or very large) number of modes, and we know that they all mix during the evolution, since we have non-vanishing Bogolyubov coefficients between the \textit{in} and \textit{out} regions. 

We will focus on the tunable transmission line analogue of Sec. \ref{sec:nation}, where far enough from the effective horizon the propagation speed of the field is constant. In this region, we have a free theory and the field modes do not interact with each other. Although the coupling to an environment could in principle introduce mode-mixing terms, the Born-Markov approximation assumes that bath correlations decay on timescales much shorter than those of the system \cite{petruccione}. This means that any and all photons lost by any mode of the field will not induce any additional interaction with the other modes, since they will not be present long enough to do so. Consequently, we can safely say that dissipative processes will affect modes individually, as if each were coupled to its own transmission line independent from the others. We can thus use Eq. \eqref{eq:lindblad} for each mode with its respective creation and annihilation operators.

For consistency, one additional aspect must be considered. In Hawking's  original derivation \cite{hawkingnt1971}, the emitted quanta originate in the immediate vicinity of the horizon. Since this region corresponds to the effective curved spacetime, the Lindblad description introduced above is not directly applicable  there. Instead, it only becomes justified once the excitation reaches the asymptotic region where the propagation speed is approximately constant and a particle interpretation can be unambiguously defined.

 We would like to state that this transit time is negligible compared with the relevant dissipation and decoherence timescales. However, as the excitations are created arbitrarily close to the  black hole horizon, this assumption requires quantitative justification.

To estimate the relevant timescale, we  numerically simulate the following semiclassical scenario: we solve the equation of motion derived for the field $\Phi$ from Eq. \eqref{eq:LagrangianoNation} using the velocity profile of Eq. \eqref{eq:perfil}. We consider an initial pulse located close to the horizon and propagating away from it. In order to obtain a conservative estimate, the pulse is initially placed at a distance of order $a$ from the horizon. Our goal is to determine the characteristic time required for the perturbation to reach the asymptotically flat region.

\begin{figure} [h!]
    \centering
    \begin{subfigure}{\linewidth}
        \centering
        \includegraphics[width=\linewidth]{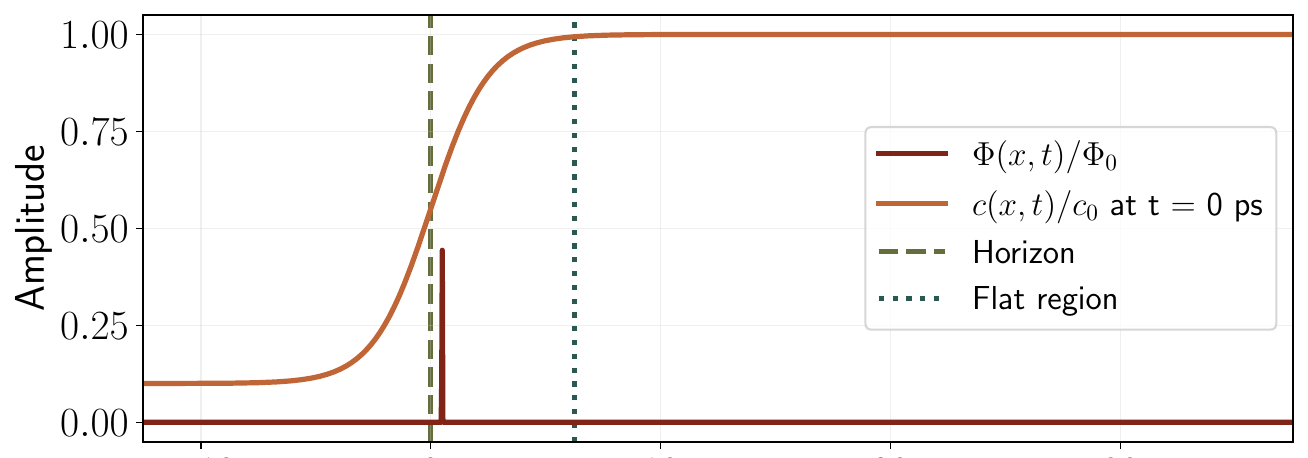}
    \end{subfigure}
    ~
    \begin{subfigure}{\linewidth}
        \centering
        \includegraphics[width=\linewidth]{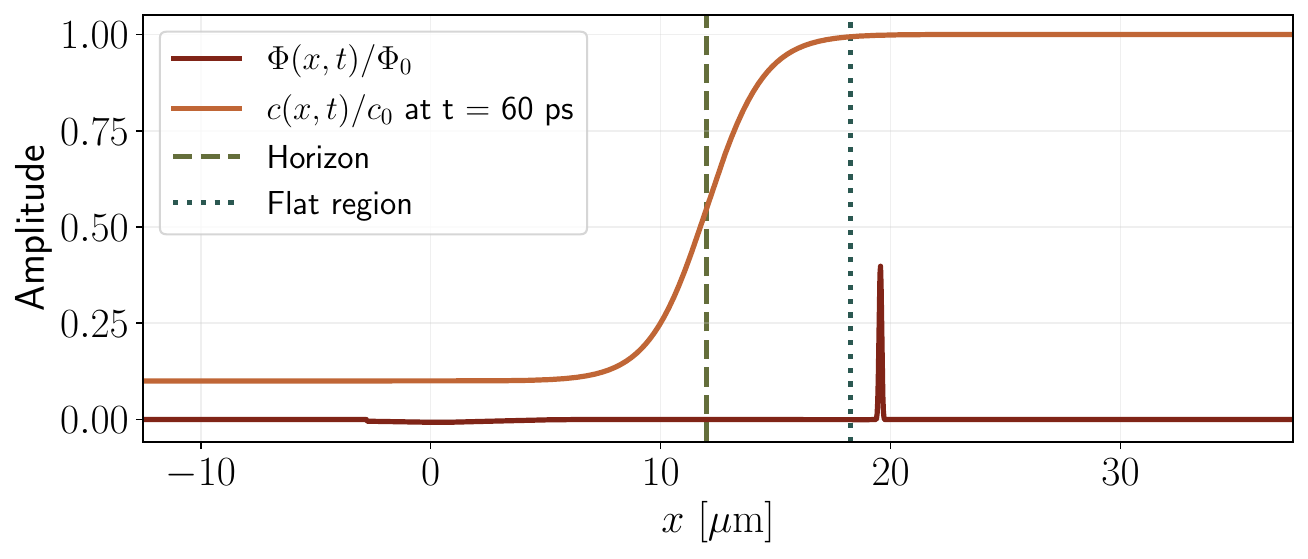}
    \end{subfigure}
    \caption{Evolution of the classical field of Eq. \eqref{eq:LagrangianoNation} (red) with the propagation velocity given by Eq. \eqref{eq:perfil} (orange). On the upper panel, the initial state of the field represents a perturbation created at a short distance $2a$ away from the horizon (green dashed). In the lower panel, after evolving for 60 ps the perturbation has escaped to the flat region, delimited by the blue dotted line.}
    \label{fig:simulaciones}
\end{figure}

Figure \ref{fig:simulaciones} shows an initial pulse at distance $2a$ from the horizon, representing an excitation of the field $\Phi$ that has just been created. For the pulse, velocities such that $u_0 = 0.55c_0$ and $v=0.45c_0$ were chosen, with $\lambda_c = 10a$. If we let the system evolve in the laboratory frame, after a time of approximately 100 ps we can observe that this hypothetical particle has escaped the curved region of spacetime. Since typical dissipation and decoherence timescales are of a few tens of $\mu$s, this propagation time is many orders of magnitude shorter. We therefore conclude that environmental effects can be safely neglected while the excitation traverses the near-horizon region, and that dissipation may be modeled only after it reaches the asymptotically flat section of the transmission line.

We, herein, make two assumptions. First, the different field modes remain effectively independent even in the presence of the environment. Second, dissipative effects are assumed to be negligible during the escape time from the near-horizon region and become relevant once the excitations reach the asymptotically flat region. Under these assumptions, we can simulate the evolution of the thermal spectrum of Eq. \eqref{eq:thermal_spectrum} and assess whether the  analogue Hawking radiation remains experimentally observable in the presence of dissipation and other imperfections.

\subsection{Effects of dissipation on the thermal spectrum}

Having formulated the theoretical framework for including dissipation, we now consider a possible experimental implementation. The transmission line of Fig. \ref{fig:circuitoNation} is coupled at its output to a measuring system, such as a transmon qubit or a SQUID. These superconducting elements can act as effectively frequency-tunable two level systems, storing excitations or being used to probe the occupation of the transmission line modes \cite{blais2021}. The Hawking excitations generated near the analogue horizon propagate along the transmission line and reach the detection stage on timescales of order picoseconds. The interaction with the environment becomes relevant at the measurement stage, where the Hawking signal is read out. In this model, we neglect imperfections and losses along the transmission line itself, so that the field evolution remains effectively unitary during propagation. Dissipation effects are therefore assumed to arise from the detection apparatus, and typical coherence and relaxation parameters reported for superconducting qubits are considered in the analysis \cite{DelGrosso2025}.

In the asymptotic flat region, where the propagation speed is constant and a particle interpretation is well defined, the field Hamiltonian can be expanded in terms of creation and annihilation operators as

\begin{equation}\label{eq:H}
    H = \int d\omega \ \omega\left( a_\omega^\dagger a_\omega + \frac{1}{2} \right) = \int d\omega \ \mathcal{H}_\omega.
\end{equation}

Here, disipation is modeled through Eq. \eqref{eq:lindblad} using relaxation and dephasing  rates representative of state-of-the-art superconducting qubits. Under the assumptions discussed above, each field mode can be treated independently. The density matrix associated with a mode of frequency $\omega$ therefore obeys

\begin{equation}\label{eq:lindblad_BH}
    \dot{\rho}_\omega = -i[ \mathcal{H}_\omega, \rho_\omega] + \kappa_{\downarrow} \mathcal{D}[ a_\omega]\rho_\omega + \kappa_\uparrow  \mathcal{D}[ a_\omega^\dagger]\rho_\omega 
    + \gamma \mathcal{D}[ a_\omega^\dagger a_\omega]\rho_\omega ,
\end{equation}

\noindent where $\kappa_\uparrow = \kappa \, n_{\text{th}}$ and $\kappa_\downarrow = \kappa (n_{\text{th}} + 1)$, with $n_{\text{th}}$ the mean number of excitations from a thermal bath (the same expression as Eq. \eqref{eq:thermal_spectrum} but with the thermal bath's temperature). 

Since typical measurements for these kinds of systems take about 200 ns, we investigate whether the Hawking signal can remain detectable over timescales of a few microseconds, namely 2 $\mu$s.  Since superconducting circuits typically operate in the  1-5 GHz range, we restrict our analysis to this frequency interval. In order to quantify whether the evolved state  remains distinguishable from the  thermal state, we compute the Hilbert-Schmidt distance

\begin{equation}\label{eq:HS}
    D(\rho_1, \rho_2) = \sqrt{\text{Tr}\left[ \left( \rho_1 - \rho_2 \right)^2 \right]}
\end{equation}

\noindent between the state at time $t$ and the bath state as a complementary measure to the number of particles. This quantity provides a useful measure of distinguishability between both states. In addition, it can be experimentally estimated with fewer resources than full state tomography  and has been shown to be relatively robust against experimental imperfections \cite{HSexp}.

\begin{figure} [h!]
    \centering
    \includegraphics[width=1\linewidth]{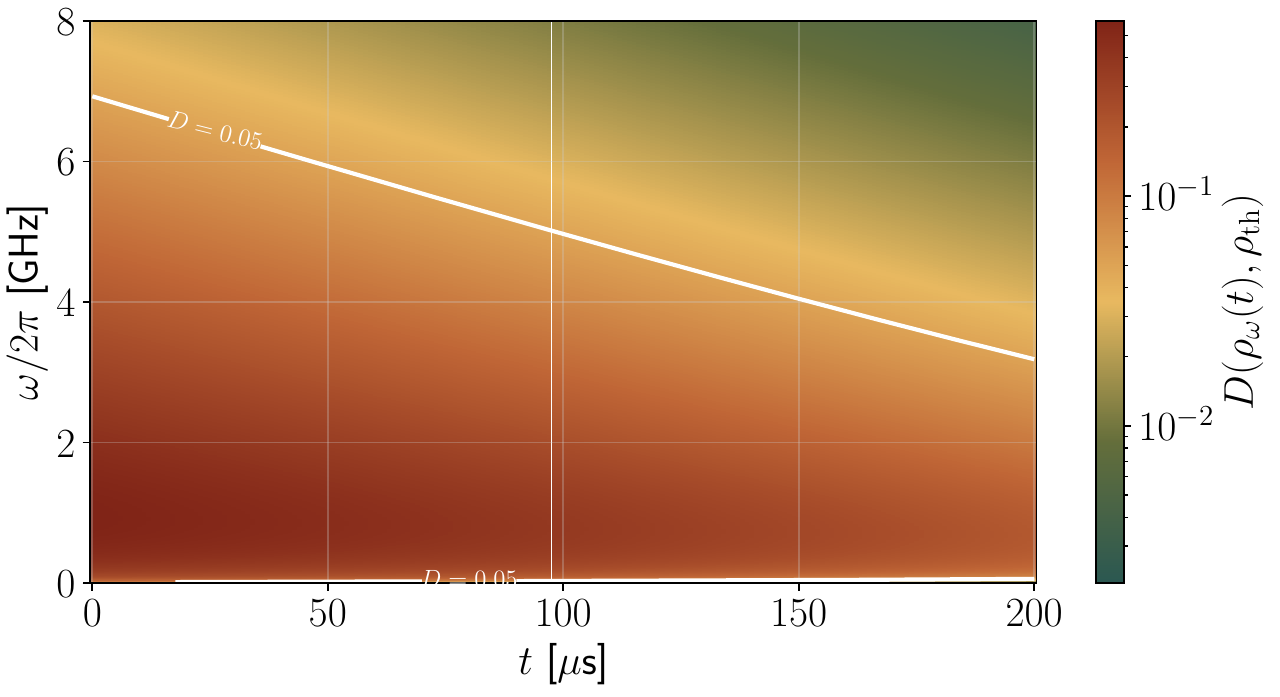}
    \caption{Hilbert-Schmidt distance between the state of the transmission line and the bath thermal state for each frequency as a function of time. The white line represents the detectability threshold.}
    \label{fig:HS_distance}
\end{figure}

 Figure \ref{fig:HS_distance} shows the Hilbert-Schmidt distance of Eq. \eqref{eq:HS} for a Hawking temperature of 100 mK, assuming dissipation and decoherence lifetimes of $\tau_\kappa = 1/\kappa = 100$ $\mu$s and $\tau_\gamma = 1/\gamma = 20$ $\mu$s respectively, and a bath at 20 mK \cite{DelGrosso2025}. For this analysis, we adopt $D=0.05$ as a practical distinguishability threshold, following Ref. \cite{HSexp}.

\begin{figure} [h!]
    \centering
    \includegraphics[width=0.95\linewidth]{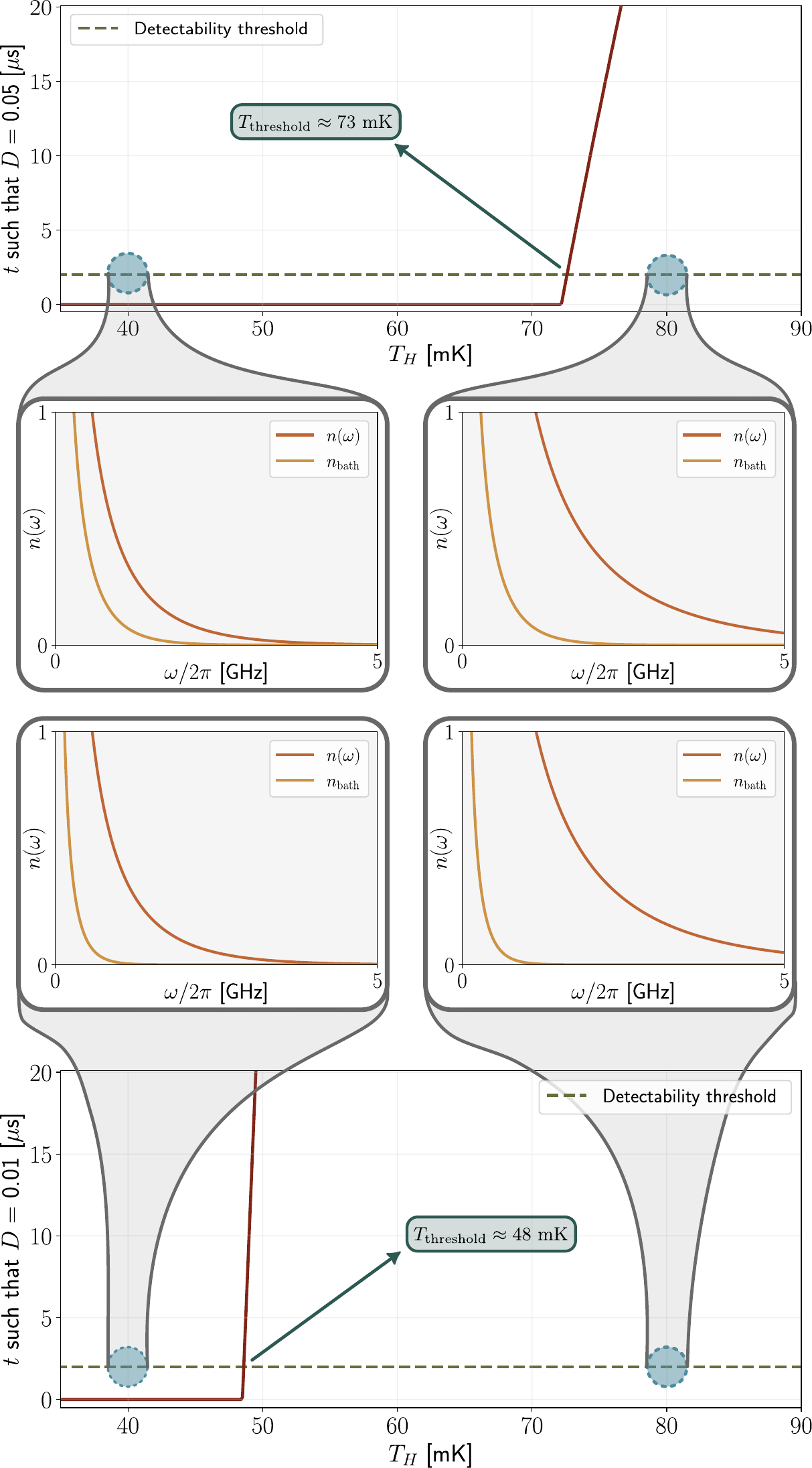}
    \caption{Time for which the radiation state becomes indistinguishable from the bath thermal state at a frequency of 5 GHz (red line). The dashed line represents the threshold time of 2 $\mu$s needed in order to measure the state. On the top, we show a setup with $\tau_\kappa = 100$ $\mu$s, $\tau_\gamma = 20$ $\mu$s, a bath at 20 mK and a threshold HS distance of 0.05; for which the minimum Hawking temperature we could realistically measure is around 73 mK. On the bottom, however, if $\tau_\kappa = 200$ $\mu$s with a reservoir at 10 mK and a threshold HS distance of 0.01, the minimum temperature reduces to 48 mK. The left and right inset axes in both panels show the particle number in the Hawking state at 40 and 80 mK respectively after the threshold time has elapsed (orange), compared to the number of excitations in the bath (yellow).}
    \label{fig:threshold}
\end{figure}

As shown in Fig. \ref{fig:HS_distance}, higher-frequency components of the Hawking spectrum become indistinguishable from the thermal bath more rapidly. For instance, the mode at  $\omega = 2\pi \times 5$ GHz reaches the threshold value $D=0.05$ earlier than the mode at $\omega = 2\pi \times 2$ GHz. Therefore, requiring the entire spectrum to remain distinguishable from the thermal background amounts to imposing that the highest-frequency mode stays above the chosen Hilbert-Schmidt threshold. By computing the time at which this condition ceases to hold for different Hawking temperatures, we determine the minimum temperature for which the Hawking signal remains distinguishable from the reservoir for 2 $\mu s$. Any  spectrum with a temperature above this value is therefore expected to persist long enough to be experimentally resolved. The resulting threshold temperature is shown in Fig. \ref{fig:threshold} for two scenarios: one using the same parameters as in Fig. \ref{fig:HS_distance} and a more optimistic case, with $\tau_\kappa = 200$ $\mu$s, a bath at 10 mK and a threshold Hilbert-Schmidt distance of 0.01.

Each point in Fig. \ref{fig:threshold} corresponds to a complete thermal Hawking spectrum evolved up to the associated time. For two representative points, we compare the resulting particle-number distribution with that of the thermal bath. As expected, when the Hawking temperature lies below the threshold value, the radiation spectrum becomes difficult to distinguish from the reservoir. For example, in the left insets of Fig. \ref{fig:threshold}, corresponding to $T_H = 40$ mK, the Hawking and bath spectra become nearly indistinguishable above approximately 3 GHz. In contrast, for temperatures above the threshold, shown in the right insets, the two spectra remain clearly distinguishable throughout the relevant frequency range.

The comparison also illustrates the impact of dissipation. For the same Hawking temperature and evolution time, the optimistic scenario exhibits significantly less degradation of the Hawking spectrum, as expected from its longer coherence times and lower bath temperature.
 
In both cases, Hawking temperatures above  73 mK remain experimentally resolvable. Since the tunable transmission line analogue discussed in Sec. \ref{sec:nation} can reach temperatures of about 113 mK, our results indicate that the analogue Hawking radiation should remain observable under either set of experimental conditions.

To ground our dissipative analysis in realistic experimental conditions, the parameters $\kappa$ and $\gamma$ are chosen to match current state-of-the-art cQED platforms \cite{blais2021, DelGrosso2025}. Single-photon relaxation lifetimes of $\tau_\kappa = 1/\kappa \sim 100\text{--}200\,\mu\text{s}$ are routinely demonstrated in modern 3D transmon architectures and low-loss superconducting waveguides, primarily limited by dielectric surface absorption. Meanwhile, a pure dephasing lifetime of $\tau_\gamma = 1/\gamma \sim 20\,\mu\text{s}$ accounts for residual $1/f$ flux noise in flux-tunable SQUID elements. Furthermore, dilution refrigerators operate stably at base temperatures of $T_{\text{bath}} = 10\text{--}20\,\text{mK}$, ensuring a minimal background thermal population in the $1\text{--}5\text{ GHz}$ frequency band. Under these conditions, the required $2\,\mu\text{s}$ detection window evaluated in Fig. \ref{fig:HS_distance} and Fig. \ref{fig:threshold} is well within the system coherence times and comfortably exceeds typical single-shot qubit readout durations ($\sim 200\text{ ns}$), confirming the experimental feasibility of the proposed protocol.

The solitonic analogue presents additional difficulties. Since there is no asymptotically flat region between the black-hole and white-hole horizons, the dissipative analysis is not as straightforward as in the tunable transmission line. Following \cite{katayama}, one possible strategy is to perform the detection in the white-hole region, where sufficiently far from the horizon the effective spacetime becomes conformally flat. Assuming comparable dissipation and decoherence timescales, the Hawking temperature of the KdV configuration above remains approximately two orders of magnitude below the experimentally resolvable range identified in our analysis. Reaching an observable signal would therefore require optimizing the transmission line specifically for the KdV case. For example, choosing $I_c = 10$ $\mu$A and $C_J=0.01$ fF yields a Hawking temperature of approximately 42 mK. Although this value remains below the threshold obtained for the tunable transmission line analogue, it could become experimentally accessible with improved detection performance beyond the optimistic parameters assumed in the present work.

Regarding the physical parameters of the transmission line architectures, the tunable dc-SQUID waveguide uses standard fabrication values ($a = 0.25\,\mu\text{m}$, $I_c = 2\,\mu\text{A}$, $C_0 = 2.87\,\text{fF}$). In contrast, for the solitonic SNAIL architecture to approach the detectable threshold (reaching $T_H \approx 42\text{ mK}$), a critical current of $I_c = 10\,\mu\text{A}$ and a junction capacitance as low as $C_J = 0.01\text{ fF}$ ($10\text{ aF}$) would be required. Such a small capacitance demands sub-micron Josephson junction areas near the current limits of electron-beam lithography, further highlighting why the tunable SQUID architecture represents a significantly more feasible path for immediate experimental implementation.

The dissipative dynamics considered in this section relies on a Born-Markov Lindblad master equation model, where photon losses (at rate $\kappa$) and pure dephasing (at rate $\gamma$) act independently on each outgoing mode once it leaves the near-horizon region. In cQED architectures, photon loss primarily arises from dielectric absorption and radiative coupling to the transmission line, whereas pure dephasing is driven by low-frequency flux noise in the SQUID loops and charge fluctuations.

We emphasize that the Markovian approximation is exceptionally well-suited for cQED platforms: at operating frequencies around $1\text{--}5\text{ GHz}$ and dilution refrigerator temperatures ($10\text{--}20\text{ mK}$), the environmental correlation time ($\tau_B \sim \hbar / k_B T \approx 0.5\text{ ns}$) is drastically shorter than the characteristic relaxation and dephasing times ($\tau_\kappa \sim 100\,\mu\text{s}$, $\tau_\gamma \sim 20\,\mu\text{s}$), as well as the detection timescale ($t \sim 2\,\mu\text{s}$). Consequently, non-Markovian memory effects are negligible over the relevant measurement window.

Alternative theoretical frameworks to treat environmental coupling include non-Ohmic spectral densities, non-Markovian baths via the Schwinger-Keldysh closed-time-path (CTP) formalism \cite{Schwinger1961, Keldysh1965, Calzetta2008}, or influence functional methods \cite{Feynman1963, Lombardo1996}. While such extensions lie beyond the scope of this work, they are expected to yield qualitatively similar detectability thresholds given the high fidelity of the Markovian picture in state-of-the-art cQED experiments.

\section{Conclusions}\label{sec:conclusiones}

We have studied two proposals for analogue black holes based in superconducting quantum circuits: a tunable dc-SQUID transmission line and  a SNAIL-based transmission line supporting solitons solutions of the Kdv and mKdV equations. We showed that both architectures generate effective horizons and analogue Hawking radiation, and we analyzed the conditions under which experimentally observable Hawking temperatures can be achieved. In the case of the tunable transmission line, we also showed that the inclusion of the physical transverse dimension of the waveguide removes the conformal ambiguity of the effective 1+1 dimensional description, allowing a consistent identification of the corresponding Painlevé-Gullstrand-Lemaître metric.

For the tunable transmission-line analogue, we identified parameter regimes that maximize the Hawking temperature while remaining consistent with the physical constraint of the circuit, obtaining temperatures as high as 113 mK. For the solitonic proposal, we analyzed the KdV and mKdV regimes, discussed parameter optimization strategies, and compared their corresponding Hawking temperatures. Our results indicate that the KdV configuration is the most promising among the solitonic analogues, although its achievable temperatures remains significantly lower than that of the tunable transmission line.

We also developed a dissipative framework based on open quantum systems techniques to assess experimental observability of the Hawking spectrum. Under a set of physically motivated assumptions, we modeled the interaction of the outgoing field modes with a thermal environment and proposed the combined use of particle-number measurements and the Hilbert-Schmidt distance as complementary indicators of distinguishability from the thermal bath. This allowed us to establish a practical detectability threshold in the presence of dissipation and thermal noise.

Our analysis shows that analogue Hawking radiation transmission line architectures are only observable above a minimum temperature threshold, which depends on the quality of the measurement setup. For the scenarios considered here, this threshold lies between approximately 48 mK and 73 mK. Consequently, the tunable transmission line proposal remains experimentally viable with currently available superconducting-circuit technology, whereas the solitonic analogue requires further optimization and more demanding experimental conditions before a clear observation of Hawking radiation can be expected.

\section*{Acknowledgments}

We thank Francisco Diego Mazzitelli, Ali Martin Zynda, Federico Guglielmucci, and Maximiliano Drelewicz for very helpful discussions. This work was supported by Consejo Nacional de Investigaciones Científicas y Técnicas (CONICET) and Universidad de Buenos Aires (UBA).

\bibliography{referencias}

\end{document}